# Elastic, electronic and magnetic properties of new oxide perovskite BaVO$_3$: a first-principles study

V.V. Bannikov[*]

*Institute of Solid State Chemistry, Ural Branch of the Russian Academy of Sciences, 620990, Ekaterinburg, Russia*

**ABSTRACT**

The structural, elastic, magnetic properties, as well as electronic structure and chemical bonding picture of new oxide 3$d^1$-perovskite BaVO$_3$, recently synthesized, were systematically investigated involving the first-principles FLAPW-GGA calculations. The obtained results are discussed in comparison with available experimental data, as well as with those obtained before for isostructural and isoelectronic SrVO$_3$ perovskite.

Keywords: oxide perovskites, first-principles calculations, elastic properties, magnetic susceptibility

## 1. Introduction

The oxide perovskites traditionally are of great material science interest due to the wide variety of their physical properties potentially useful for practical applications: for instance, ferroelectric, magnetic, electronic, transport properties, structural phase transitions, colossal magneto-resistance, ionic conductivity, catalytic activity and so on. The important feature of these compounds is the essential interplay between their chemical composition and the structural type, and, as a consequence, their physical properties, which may differ even for members of isoelectronic series. As an well known example, SrTiO$_3$ oxide perovskite possesses the cubic *Pm-3m* structure at room temperature and is a paraelectric, while BaTiO$_3$, which is isoelectronic to latter, prefers the tetragonal *P4mm* structure (due to the discrepancy of the Sr$^{2+}$ and Ba$^{2+}$ cations radii) and is a ferroelectric. Moreover, the oxide perovskites containing metal atoms with partially filled *d*-shells can exhibit a rich multiplicity of unique physical properties, due to the correlation between the spin, orbital and lattice degrees of freedom: non-collinear magnetic ordering, metal-insulator transitions, phase separation, charge ordering and so on (see review [1] and references therein). However, despite the increasing number both of experimental and theoretical investigations devoted to the properties of transition metals oxide perovskites, the common and consistent theory able to explain the variety of mentioned phenomena still remains open question. It seems reasonable to explore in details the physical properties and electronic structure of so-called $3d^1$-perovskites $A^{2+}VO_3$, where $A^{2+}$ is a divalent alkali-earth metal ion, and vanadium ion is supposed in purely ionic model to have a single electron in $3d$-shell, because these perovskites can be treated as a simplest model for better understanding the interplay between orbital and spin states, as well, as the influence of lattice distortions and electronic correlation effects.

---

[*] Corresponding author.
 *E-mail address*: bannikov@ihim.uran.ru

By now, a considerable amount of investigations devoted to the properties of $A^{2+}VO_3$ perovskites has been published. So, in early experimental paper [2] it was shown that both $SrVO_3$ and $CaVO_3$ are metallic, and the magnetic properties of $SrVO_3$ can be satisfactorily explained within free-electron model, assuming the electron correlation to be small, however, $SrVO_3$ is a cubic perovskite, while $CaVO_3$ is orthorhombic or tetragonal, depending on the synthesis conditions. In [3,4] the elastic and thermoelectric properties of $SrVO_{3-\delta}$ ceramics were experimentally investigated, indicating relatively high negative Seebeck coefficient and thermal expansion coefficient and discussing its potential application as $n$-type oxide thermoelectric material and solid oxide fuel cell. The paper [5] reports the successful synthesis of high quality stoichiometric $SrVO_3$ thin films and offers their potential use as a growth template for the epitaxial integration of functional oxide materials with perovskite-like structure. The first principles calculations of $SrVO_3$ ground state band structure were performed in [6-8]. The crystal structure, magnetic properties and electrical conductivity of $CaVO_3$ perovskite were experimentally investigated in [9], according to this data, it crystallizes in orthorhombic *Pbnm* structure, exhibits predominantly Pauli-paramagnetic character of magnetic susceptibility and metallic behavior between 2 and 300 K with a positive magneto-resistance at low temperatures, the same properties were observed in epitaxial $CaVO_3$ thin films [10]. However, it is well-known, that $SrVO_3$ and $CaVO_3$, being simple metals and Pauli paramagnets in their ground state, behave as systems with strong electronic correlation in spectroscopic investigations (see, for instance, [11] and references therein), moreover, the correct interpretation of spectroscopic data using first-principles calculations implies the on-site Coulomb repulsion energy $U$ to be frequency dependent, or different for various electronic transitions [12].

Till recently, $BaVO_3$ compound was known to exist in hexagonal modification only [13], and no experimental evidences of its perovskite phase were available. In the recent paper [14] K. Nishimura and co-workers report the successful high-pressure synthesis of $BaVO_3$ with cubic *Pm-3m* perovskite structure observed from room temperature down to 23 K, revealing no ferroelectric structural transitions, similar to those in $BaTiO_3$. The measurements of electrical resistivity and magnetic susceptibility have shown, that $BaVO_3$, like $SrVO_3$, possesses metallic conductivity with $\rho(T)=\rho_0+A \cdot T^2$, and is a Pauli paramagnet with $\chi=1.3 \cdot 10^{-4}$ $cm^3/mol$ above 20 K, being characterized in this way as a Fermi liquid metal. As far, as author knows, the only paper devoted to the first-principles investigation of structural and electronic properties of $BaVO_3$ perovskite is [8], where Shu-yao Yan et al have predicted the absence of ferroelectric instability for it, however, the results on band structure both of $SrVO_3$ and $BaVO_3$ presented there seem to be disputable (see details in Discussion). The data concerning the elastic properties of $BaVO_3$ perovskite phase are absent for today.

In present paper, based on first-principles FLAPW-GGA band structure calculations, the elastic properties of $BaVO_3$ perovskite phase were systematically explored for the first time both for monocrystals and the polycrystalline species, the electronic, magnetic properties and the picture of chemical bonding also were investigated and discussed in comparison with available experimental data and results reported before for isoelectronic SrVO$_3$ perovskite.

## 2. Computational aspects

In accordance with [14], BaVO$_3$ is treated as an ideal cubic perovskite (space group *Pm-3m*, Z=1), the atomic positions are O: 3*c* (½,½,0), Ba: 1*a* (0,0,0), and V: 1*b* (½,½,½). The spin-polarized band structure calculations were carried out by means of the full-potential method of linearized augmented plane waves with mixed basis APW+lo (FLAPW) implemented in the WIEN2k suite of programs [15]. The generalized gradient approximation (GGA) to exchange-correlation potential in PBE form [16] was used. The plane-wave expansion with $R_{MT} \times K_{MAX}$ equal to 8, and the sampling with 120 *k*-points in the irreducible part of the Brillouin zone (BZ), corresponding to 15×15×15 *k*-mesh, were used. The energy cut-off separating the core and valence states was chosen to be –6.0 Ry. The linearization energy $E_l$ for V-3*d* states was specified to be 0.5 Ry. The *muffin-tin* (MT) radii were chosen equal to 2.3 a.u. for Ba, 2.0 a.u. for vanadium and 1.5 a.u. for oxygen atoms. The self-consistent calculations were considered to have converged, when the differences in the total energy of the crystal and in the electronic charge inside MT spheres did not exceed 0.01 mRy and 0.001*e*, respectively, as calculated at consecutive steps. To calculate the density of electronic states (DOS) the modified tetrahedron method [17] was used, and magnetic moments, both atomic (inside the corresponding MT spheres) and per total cell, were computed by integration over the partial and total DOSs, respectively.

## 3. Results and discussion

*Structural and elastic properties*

As an initial step, the volume of BaVO$_3$ perovskite cubic cell was optimized, and the equilibrium lattice constant was found to be $a_0$ = 3.953 Å, being in reasonable agreement with experimental value 3.943 Å [14]. The fitting of $U(V)$ curve (energy per cubic unit cell *vs* cell volume) with Murnaghan equation:

$$U(V) = U_0 + (B_0 \cdot V/B`) \cdot [(V_0/V)^{B`}/(B`-1)+1] - B_0 \cdot V_0/(B`-1) \quad (1)$$

gives equilibrium unit cell volume to be $V_0$ = 417.44 bohr$^3$, and bulk modulus $B_0$ and its pressure derivative $B`$ at zero pressure to be 173.86 GPa and 2.64, respectively. It is known, that $B`$ may be considered as a measure of crystal anharmonicity with respect to hydrostatic pressure. To characterize it quantitatively, let us take into account the $P(V)$ Murnaghan equation in the form $P(\delta) = (B_0/B`) \cdot [(1+\delta)^{-B`}-1]$, where $\delta=(V-V_0)/V_0$, and the linear term of its expansion in a series, $P_{\text{lin}}(\delta)=-B_0 \cdot \delta$, which simply corresponds to the Hooke`s law (if higher order terms are ignored), then introduce the quantity:

$$\Delta(\delta) = 1 - P(\delta)/P_{\text{lin}}(\delta) \quad (2)$$

which represents the relative deviation from the Hooke`s law at given compressive strain $\delta$. It can be shown that at $|\delta|<0.1$ and reasonable $B`$ values (1÷5) the simple relation $\Delta(\delta) \approx (B`+1) \cdot \delta/2$ takes place, or, in other words, at compressive strains $|\delta|<2 \cdot |\Delta|/(B`+1)$ the deviation does not exceed $|\Delta|$. The corresponding hydrostatic pressure value may be taken from Murnaghan $P(\delta)$ equation. So, for BaVO$_3$ it was estimated, that the relative deviation from the Hooke`s law does not exceed 1%, 5% and 10% at hydrostatic pressure values lower than 0.94 GPa, 4.55 GPa and 8.68 GPa, respectively.

To calculate three independent elastic constants, $C_{11}$, $C_{12}$ and $C_{44}$, the following deformations (each is characterized with single parameter) applied to the cubic structure were modeled: hydrostatic compression (strain tensor $e_{xx}=e_{yy}=e_{zz}=\delta/3$, $e_{xy}=e_{xz}=e_{yz}=0$), tetragonal volume-conserving distortion

($e_{xx}=e_{yy}= -e_t/3$, $e_{zz}=2 \cdot e_t/3$, $e_{xy}=e_{xz}=e_{yz}=0$, where $e_t = c/a-1$), and rhombohedral distortion along the body diagonal of the cell ($e_{xx}=e_{yy}=e_{zz}=\gamma/3$, $e_{xy}=e_{xz}=e_{yz}=2 \cdot \gamma/3$, where $\gamma$ is relative variation of body diagonal), the fitting of calculated unit cell energy *vs* corresponding parameter dependence with second-order polynomial allows to determine the values of $C_{11}+2 \cdot C_{12}$, $C_{11}-C_{12}$, and $C_{11}+2 \cdot C_{12}+4 \cdot C_{44}$ linear combinations, respectively. The obtained values of elastic constants for $BaVO_3$ are listed in Table 1 in comparison with the ones for $SrVO_3$ [7], calculated within the similar computational scheme, it is seen, that for both oxide perovskites the well-known criteria of mechanical stability for cubic crystals ($C_{11}+2 \cdot C_{12}>0$, $C_{11}-C_{12}>0$, $C_{44}>0$) is satisfied. It should be noted, though, that this criteria does not take into account the possible relative displacement of various sub-lattices of complex compounds, in particular, the soft polar mode in $ABO_3$ perovskites (the opposite moving of B atoms against the oxygen sublattice), which may result in ferroelectric instability. Nevertheless, it was predicted in [8], that $BaVO_3$ perovskite is stable with respect to ferroelectric soft mode, and experimentally confirmed in [14] by X-ray diffraction measurements, finding out no symmetry lowering in $(23 \div 300)$K temperature range.

Let us compare the calculated elastic properties of $BaVO_3$ and $SrVO_3$ monocrystals summarized in Table 1. The bulk modulus ($B_0$) of $BaVO_3$ is ~5% less than one of $SrVO_3$, in accordance with the well-known simple trend, $B_0 \sim 1/V_0$, taking place for the series of crystals with the same structure and similar chemical bonding picture (see below), respectively, the compressibility ($\beta$) of $SrVO_3$ is lower than that of $BaVO_3$. On the contrary, the shear modulus ($G=C_{44}$) and tetragonal shear modulus ($G_t$) of $BaVO_3$ are ~10% and ~28%, respectively, greater than these of $SrVO_3$, indicating that $BaVO_3$ is stiffer with respect to any low-symmetry deformations. It is interesting to note, that the Cauchy pressure (*CP*), predicted for $SrVO_3$ to be positive, for isoelectronic $BaVO_3$ adopts the negative value, as well, as for semiconducting $SrTiO_3$ and

SrZrO$_3$ perovskites [7]. It was proposed by Pettifor [18], that the compounds with positive *CP* are characterized with predominantly metallic chemical bonding, while the covalent contribution is deciding, if *CP* is negative. So, it may be expected, that the elastic properties of BaVO$_3$ should coincide to these of predominantly covalent compound, as distinct from predominantly metallic SrVO$_3$. The dependence of Young`s modulus (*Y*) on the selected direction ($\boldsymbol{n}=(n_x, n_y, n_z)$, $|\boldsymbol{n}|=1$) in cubic crystal is expressed, as follows:

$$Y(\boldsymbol{n}) = [(C_{11}+C_{12})/\{(C_{11}+2 \cdot C_{12}) \cdot (C_{11}-C_{12})\} + $$
$$+ (1-A_Z) \cdot (n_x^2 \cdot n_y^2 + n_x^2 \cdot n_z^2 + n_y^2 \cdot n_z^2)/C_{44}]^{-1} \qquad (3)$$

where $A_Z = 2 \cdot C_{44}/(C_{11} - C_{12})$ is so-called Zener anisotropy index, if $A_Z=1$, *Y* does not depend on the direction $\boldsymbol{n}$, while if $A_Z>1$, the Young`s modulus is maximal in [111] and minimal in [100] direction, and *vica versa*, if $A_Z<1$. As is seen from Table 1, both for BaVO$_3$ and SrVO$_3$ $A_Z>1$, and maximal value of *Y* corresponds to [111] direction, at the same time, for BaVO$_3$ the values of *Y* in all specified directions are greater than those for SrVO$_3$, and *Y*($\boldsymbol{n}$) "dispersion" for former is weaker than for latter (taking into account the discrepancy between maximal and minimal *Y* values in [111] and [100] directions, respectively). The corresponding values of Poisson`s ratio (ν) were calculated as $\nu(\boldsymbol{n})=[1-Y(\boldsymbol{n})/(3 \cdot B_0)]/2$, and for BaVO$_3$ they are smaller than for SrVO$_3$ in each specified direction. It is commonly known, that for typical metallic compounds (aluminum and copper alloys, for instance) ν~0.3, while for compounds with predominant covalent bonding the Poisson`s ratio value is smaller (~0.1÷0.2), so this result is in accordance with assumption about the stronger covalent bonding component in BaVO$_3$ in comparison with SrVO$_3$.

Knowing the values of elastic constants for monocrystals, it is possible to estimate the elastic characteristics of the corresponding polycrystalline species, using the Voigt-Reuss-Hill (VRH) approximation [19]. For cubic crystals, the polycrystalline bulk modulus both in Voigt and Reuss approaches equals to that of monocrystals, the polycrystalline shear modulus in Voigt and

Reuss approximations is $G_V=(C_{11}-C_{12}+3 \cdot C_{44})/5$ and $G_R=5 \cdot (C_{11}-C_{12})/[4+3 \cdot (C_{11}-C_{12})/C_{44}]$, respectively, and finally, in VRH approach it is taken as arithmetic mean $G_{VRH} = (G_V+G_R)/2$. The calculated values of bulk and shear moduli, Young`s modulus $Y_{VRH}=(9 \cdot B_0)/(1+3 \cdot B_0/G_{VRH})$ and Poisson`s ratio $\nu_{VRH}=[1-Y_{VRH}/(3 \cdot B_0)]/2$ for polycrystalline BaVO$_3$ and SrVO$_3$ perovskites are listed in Table 1, the results on shear and Young`s moduli for SrVO$_3$ polycrystals are in reasonable agreement with available experimental values for non-stoichiometric SrVO$_{3-\delta}$ ceramics (90 GPa and 224 GPa, respectively) [3], for BaVO$_3$ perovskite the experimental data are absent yet. Generally, the results for polycrystalline species confirm the conclusion mentioned above, that BaVO$_3$ is stiffer and more covalent than SrVO$_3$ ($G_{VRH}$ and $Y_{VRH}$ for former are greater than for latter, and *vica versa* for $\nu_{VRH}$). The $G/B_0$ ratio acts as an indicator of Pugh`s criteria [20] of ductile/brittle behavior: if $G/B_0>0.571$, the substance behaves in a brittle manner, if not, it is ductile. As is seen from Table 1, BaVO$_3$ is expected to be brittle, while SrVO$_3$ should be ductile, and this is consistent with proposed more covalent character of inter-atomic bonding in BaVO$_3$.

*Electronic and magnetic properties*

As it would be expected, the calculated band structure of BaVO$_3$ in many features is similar to that of SrVO$_3$ described in details in [6,7], so it will be concerned in brief. The total and partial DOS for BaVO$_3$ obtained within spin-polarized FLAPW–GGA band structure calculations are presented in Fig.1. It is seen, that band structure of BaVO$_3$ consists in completely filled valence band (about 2.0 – 6.5 eV below $E_F$) and partially filled conduction band separated from the former by gap ~ 0.92 eV, in this way, BaVO$_3$ possesses metallic conductivity, in accordance with experiment [14]. The valence band is composed predominantly of O-2$p$ states with admixture both of barium (the peak in the middle of band) and vanadium states (V-3$d$(e$_g$) in the bottom of band and V-3$d$(t$_{2g}$) in its middle), while the conduction band is composed mainly of V-3$d$(t$_{2g}$) states with only small admixture of O-2$p$ states, the sharp peak of V-3$d$(t$_{2g}$) states is located ~ 0.8 eV above E$_F$, and the top of this band intersects with the bottom of vacant V-3$d$(e$_g$) band at ~1.05 eV above E$_F$. The calculated magnetic moments of vanadium atom and per total unit cell are small: 0.039 µ$_B$ and 0.059 µ$_B$, respectively (the moments of other atoms are negligible), and the difference of "spin up/down" DOS values at Fermi level is insignificant: N↑(E$_F$)=0.939 states/(eV·cell) and N↓(E$_F$)=0.884 states/(eV·cell), corresponding to spin polarization |N↑ − N↓|/(N↑+N↓) ~ 3%, at the same time, the partially filled conduction band suffers in BaVO$_3$ small, but perceptible spin splitting Δ$_s$ ~ 0.054 eV (see Fig.1 and the magnified DOS plot there).

It is clear from $E(\mathbf{k})$ band structure depicted in Fig.1, that only high-dispersive bands for specified directions of the BZ are present in the vicinity of E$_F$ corresponding to T<10$^3$ K. In the rough, the band structure of SrVO$_3$ and BaVO$_3$ perovskites (related to their ground state) may be regarded within the simple rigid-band model – as that of semiconducting SrTiO$_3$ and BaTiO$_3$ (cubic modification), respectively, with one additional electron partially filling the conduction band, initially empty (see [7] and Fig.2 therein). At the same time, the quasi-flat $E(\mathbf{k})$ bands exist for X-M direction of BZ (at ~0.8 eV above E$_F$), corresponding to the vacant sharp peak of V-3$d$(t$_{2g}$) DOS, and for Γ-X direction, at ~0.8 eV below E$_F$ and ~1.0 eV above E$_F$, corresponding to the floors of partially filled V-3$d$(t$_{2g}$) and vacant V-3$d$(e$_g$) bands, respectively, in this way, the essential role of electronic correlation effects may appear in BaVO$_3$ due to the optical excitations of near-Fermi electrons to vacant states of quasi-flat bands. Note, however, that the attempt to take the electronic correlation in BaVO$_3$ into account formally within GGA+$U$ approach [21], using the unified value of $U_{eff} = U - J_H$ (where $J_H$ is the energy of Hund exchange interaction) for all V-3$d$ states, results (at $U_{eff}$ >0.7 eV) in ground state electronic structure, characterized with considerable magnetic moments

($\sim$1 $\mu_B$) of vanadium atoms and 100% spin polarization of DOS at $E_F$, and that is inconsistent with experimental data [14], establishing the Pauli paramagnet-like behavior of BaVO$_3$. The correct modeling of optical properties both of SrVO$_3$ and BaVO$_3$, in other words, the correct simultaneous reproducing both excited states and the ground one in band structure calculations, is not a trivial problem and requires the special approaches, for instance, using the frequency-dependent value of $U$ energy (see [12] for details), which are, though, far beyond the plot of this paper.

The Fermi surface (FS) obtained for BaVO$_3$ is presented in Fig.2 (shown for "spin up" electronic sub-system only, because the discrepancy between FS topology for "spin up/down" sub-systems is small). It is quite similar to the FS of isoelectronic CaVO$_3$ perovskite (cubic modification) reported in [11] and consists of three sheets: two closed surfaces (1,2 in Fig.2) with the center at $\Gamma$ point and the "crossed tubes"-like surface (3 in Fig.2) oriented along $\Gamma$-X directions. Obviously, the closed sheets 1 and 2 are related to the "electron pockets" with the minimum at $\Gamma$-point (see $E(\boldsymbol{k})$ band structure in Fig.1, where the almost degenerate bands in R-$\Gamma$-X and $\Gamma$-M directions are shown), while the open sheet 3 may be attributed to the "hole pockets" with the maxima at R- and M-points (bands in R-$\Gamma$ and M-$\Gamma$ directions in Fig.1). All the $\boldsymbol{k}$-states related to near-Fermi bands inside the "crossed tubes" sheet are occupied, while the $\boldsymbol{k}$-states outside it are vacant (at T=0 K).

To characterize the chemical bonding picture in BaVO$_3$ the maps of electronic charge density distribution for (110) and (100) crystallographic planes are presented in Fig.3 (maps 1 and 2, respectively). It is seen, that strong direct bonds between V and O atoms take place, so the covalent contribution to the chemical bonding in BaVO$_3$ perovskite is essential. Actually, the effective atomic charges, estimated in the frameworks of Bader`s scheme [22] are: $Q$(Ba)=+1.55 $e$, $Q$(V)=+2.02 $e$ and $Q$(O)= −1.19 $e$, this considerably differs from the values assumed in purely ionic model ($Ba^{2+}V^{4+}O_3^{2-}$), indicating that the chemical bonds in $BaVO_3$, as well, as in $SrVO_3$ [7], are of mixed covalent-ionic character, so the term "$3d^1$-perovsiktes", as a matter of fact, is conventional. The main distinctive feature of bonding picture in $BaVO_3$ is the presence of additional covalent Ba-O bonds (map 2 in Fig.3), while the directional Sr-O bonds in $SrVO_3$ are absent (see [7] and Fig.4 therein), it is in agreement with the preceding conclusion, that inter-atomic bonding in $BaVO_3$ perovskite should be of more covalent character, than that is in $SrVO_3$. Note also, that the electrons of the conduction band do not participate in the formation of directional inter-atomic bonds in $BaVO_3$, as it is seen in map 3 (Fig.3), where the partial contribution of conduction band occupied states to charge density in (200) plane is depicted.

As mentioned above, both atomic and total cell magnetic moments obtained from band structure calculations are quite small, so $BaVO_3$ may be treated as a Pauli paramagnet, in accordance with experimental data on temperature dependence of magnetic susceptibility $\chi(T)$ measurements [14]. However, the formal estimation of the molar paramagnetic susceptibility as $\chi_P = 2 \cdot \mu_B^2 \cdot N(E_F) \cdot N_A / 3$ (where $N_A$ is Avogadro`s number, $N(E_F) \approx N_\uparrow(E_F) + N_\downarrow(E_F)$, and the factor 2/3 comes from taking into account the Landau diamagnetic contribution) gives for $BaVO_3$ the value $\chi_P = 3.93 \cdot 10^{-5}$ cm$^3$/mol, this is about three times less than the experimentally determined $\chi_P = 1.3 \cdot 10^{-4}$ cm$^3$/mol [14]. The similar situation takes place for $SrVO_3$, for which $N(E_F) = 1.68$ states/(eV·cell) [7] and, respectively, $\chi_P = 3.6 \cdot 10^{-5}$ cm$^3$/mol, but experimental temperature-independent contribution to $\chi$ was found to be about $1.5 \cdot 10^{-4}$ cm$^3$/mol [2]. With no purpose to analyze the contributions of different origins to $\chi$ of $BaVO_3$ in details, it is possible, nevertheless, to give a simple explanation of this discrepancy. For an ideal Pauli paramagnet the spin polarization of bands is absent in zero field, however, in $BaVO_3$ the partially filled conduction band suffers the spin splitting $\Delta_s$ (Fig.1). To explain it, the influence of exchange interactions between the electrons in the conduction band should be taken into account, which is described in the simplest way by effective Weiss molecular field $B_W=\lambda \cdot J$, where $J$ is the magnetization, and $\lambda$ is the molecular field constant. The molecular field $B_W$ results in the "exchange-enhanced" effective value of Pauli paramagnetic susceptibility $\chi_{eff} = \chi_P \cdot (1-\lambda \cdot \chi_P)^{-1}$ [23, 24]. Since the spin splitting $\Delta_s = 2 \cdot \mu_B \cdot \lambda \cdot J_s$ in zero external field, where $J_s$ is the spontaneous magnetization at T=0 K, supposed to be simply proportional to total cell magnetic moment (denoted as $\eta \cdot \mu_B$), it is easy to derive the relation:

$$\chi_{eff} = \chi_P \cdot [1 - \Delta_s \cdot N(E_F)/(3 \cdot \eta)]^{-1}, \qquad (4)$$

that yields for $BaVO_3$ the value of effective Pauli paramagnetic susceptibility $\chi_{eff}=8.85 \cdot 10^{-5}$ cm$^3$/mol, which is considerably closer to the experimental one. Certainly, the description in this rough model cannot ensure the accurate interpretation of experimental data, nevertheless, it is able to explain qualitatively, why so large discrepancies between experimentally determined $\chi_P$ and its formal estimations within first-principles calculations may occur. Further, the Pauli susceptibility $\chi_P$ slightly depends on T, including the small contribution $\sim (k_B T)^2$ and higher orders terms dependent on DOS behavior at $E_F$ vicinity [23], however, the more detailed experimental data are required to compare with first-principles estimations of temperature effects.

In conclusion, let us concern the results on $BaVO_3$ electronic structure obtained before by Shu-yao Yan et al [8]. According to their paper, both $SrVO_3$ and $BaVO_3$ perovskites are characterized with 100% spin polarization of near-Fermi electronic states and considerable magnetic moments (~2.2–2.6 $\mu_B$) of V atoms, however, it is inconsistent with experiment [2, 14] revealing the Pauli paramagnet-like behavior for both oxide perovskites. Most likely, the reason of this disagreement is simply insufficiently dense **k**-mesh sampling (4×4×4) over the BZ used by authors of [8] in the calculations. To clarify this viewpoint, the calculations of $BaVO_3$ band structure and its derivative values (energy per unit cell, $N_{\uparrow,\downarrow}(E_F)$ and magnetic moments of vanadium atoms) were performed for equilibrium cell geometry at *k*-mesh samplings with different thickness, the results are summarized in Table 2. It is seen, that too sparse *k*-mesh sampling results in essential spin polarization of near-Fermi states and considerable magnetic moments of V atoms, which, however, decrease to small, more or less constant values, as *k*-mesh becomes thicker, though the unit cell energy is not very sensitive to that. According to Table 2, it is most likely that 15×15×15 *k*-mesh is quite enough to reproduce the band structure of $BaVO_3$ correctly.

## 4. Summary

The systematical theoretical study of structural, elastic, electronic and magnetic properties of recently synthesized $BaVO_3$ oxide perovskite in comparison with isoelectronic $SrVO_3$ has been performed. It was shown, that $BaVO_3$ is expected to be more compressible, but stiffer with respect to shear and other low-symmetry deformations, than $SrVO_3$ is, the estimated maximal value of Young`s modulus for it is ~303 GPa and corresponds to [111] direction in crystal. In contrast to Sr-based perovskite, $BaVO_3$ is characterized with negative Cauchy pressure, revealing that the covalent contribution to chemical boning is more considerable for it . The brittle behavior is expected for polycrystalline $BaVO_3$, while $SrVO_3$ should be ductile. The features of $BaVO_3$ band structure, Fermi surface topology and chemical bonding picture have been discussed. Comparing the results of band structure calculations with $\chi(T)$ experimental data, it was proposed, that $BaVO_3$ behaves in magnetic field as "exchange-enhanced" Pauli paramagnet.

**Table 1.** The calculated unit cell parameters ($a_0$), elastic constants ($C_{11}$, $C_{12}$, $C_{44}$), bulk modulus ($B_0$), compressibility ($\beta$), tetragonal shear modulus ($G_t$), Cauchy pressure (*CP*), Zener anisotropy index ($A_Z$), Young`s modulus ($Y_{[100]}$, $Y_{[110]}$, $Y_{[111]}$) and Poisson`s ratio ($\nu_{[100]}$, $\nu_{[110]}$, $\nu_{[111]}$) in specified directions, polycrystalline shear modulus in Voight ($G_V$), Reuss ($G_R$) approximations, and averaged one ($G_{VRH}$), polycrystalline Young`s modulus ($Y_{VRH}$) and Poisson`s ratio ($\nu_{VRH}$), and Pugh indicator ($G/B_0$) for BaVO$_3$ and SrVO$_3$ cubic perovskites.

| Parameter | BaVO$_3$ | SrVO$_3$ |
|---|---|---|
| $a_0$, Å | 3.953 | 3.866* |
| $C_{11}$, GPa | 286.71 | 269.88* |
| $C_{12}$, GPa | 116.46 | 137.35* |
| $C_{44}$, GPa | 125.61 | 113.63* |
| $B_0 = (C_{11}+2 \cdot C_{12})/3$, GPa | 173.210 | 181.527 |
| $\beta = 1/B_0$, GPa$^{-1}$ | 0.00577 | 0.00550 |
| $G_t = (C_{11} - C_{12})/2$, GPa | 85.125 | 66.265 |
| $CP = (C_{12} - C_{44})$, GPa | -9.150 | 23.720 |
| $A_Z = 2 \cdot C_{44}/(C_{11} - C_{12})$ | 1.47559 | 1.71478 |
| $Y_{[100]}$, GPa | 219.428 | 177.230 |
| $Y_{[110]}$, GPa | 276.953 | 245.713 |
| $Y_{[111]}$, GPa | 303.472 | 282.040 |
| $\nu_{[100]}$ | 0.28886 | 0.33728 |
| $\nu_{[110]}$ | 0.23351 | 0.27440 |
| $\nu_{[111]}$ | 0.20799 | 0.24105 |
| $G_V$, GPa | 109.416 | 94.684 |
| $G_R$, GPa | 105.534 | 88.365 |
| $G_{VRH} = (G_V + G_R)/2$, GPa | 107.475 | 91.525 |
| $Y_{VRH}$, GPa | 267.166 | 235.067 |
| $\nu_{VRH}$ | 0.24292 | 0.28417 |
| $G/B_0$ | 0.62048 | 0.50419 |

* Ref. [7],

**Table 2.** The spin-polarized DOS at Fermi level ($N_{\uparrow,\downarrow}(E_F)$), magnetic moment of vanadium atom ($\mu(V)$), and total energy per unit cell (E) calculated for $BaVO_3$ at different ***k***-mesh samplings.

| k-mesh | $N_\uparrow(E_F)$, states/eV | $N_\downarrow(E_F)$, states/eV | $\mu(V)$, $\mu_B$ | E, Rydb |
|---|---|---|---|---|
| 4×4×4 | 1.332 | 0 | 0.814 | Set to zero |
| 7×7×7 | 1.237 | 0.671 | 0.474 | 0.001307 |
| 10×10×10 | 0.921 | 0.885 | 0.030 | 0.000831 |
| 15×15×15 | 0.939 | 0.884 | 0.039 | 0.000952 |

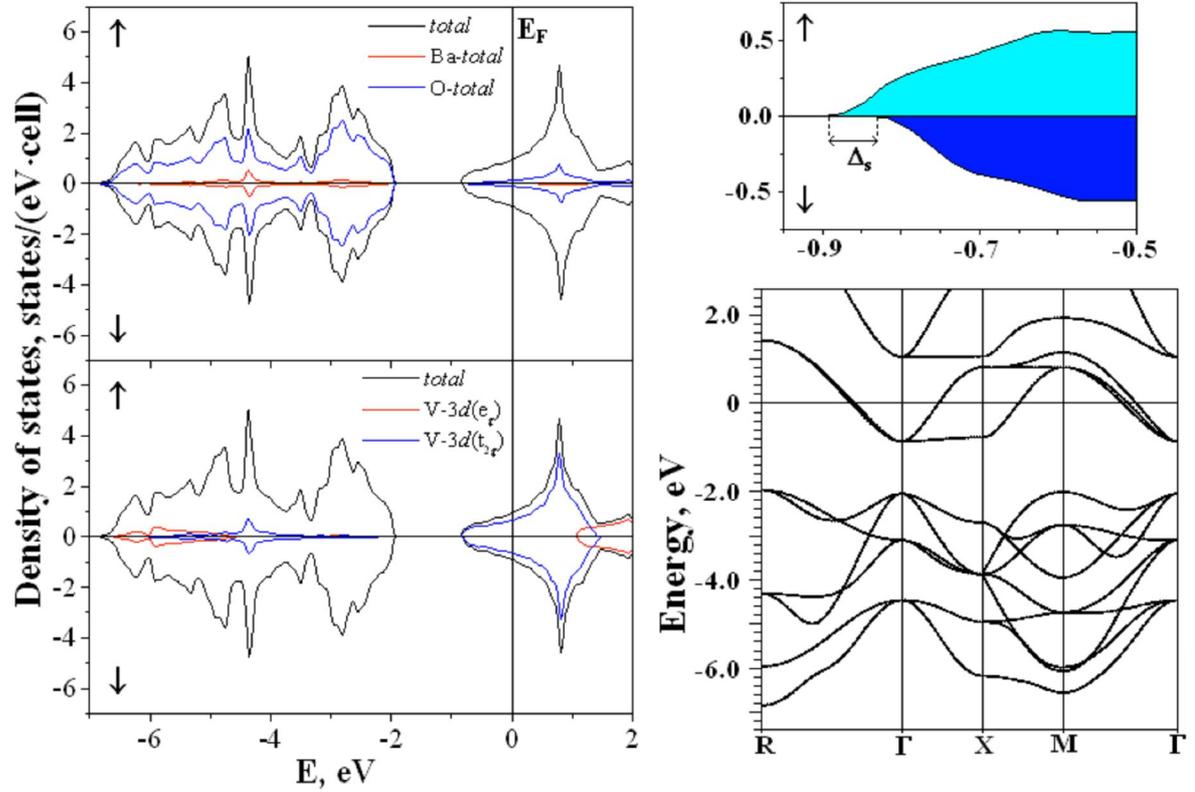

**Fig.1.** Total and partial densities of electronic states (*left*), the magnified DOS in the vicinity of the conduction band bottom (*upper right*, in the same units) and $E(\mathbf{k})$ band structure for BaVO$_3$ (*lower right*, depicted only for "spin-up" electronic sub-system, because the small spin splitting is insignificant in the context of whole band structure discussion).

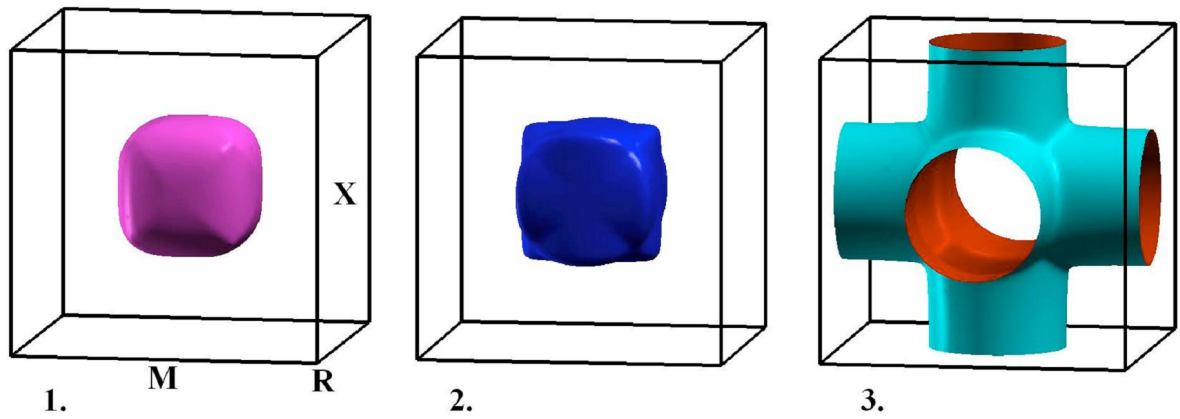

**Fig.2.** The Fermi surface calculated for $BaVO_3$ perovskite. The individual sheets composing it, are shown separately.

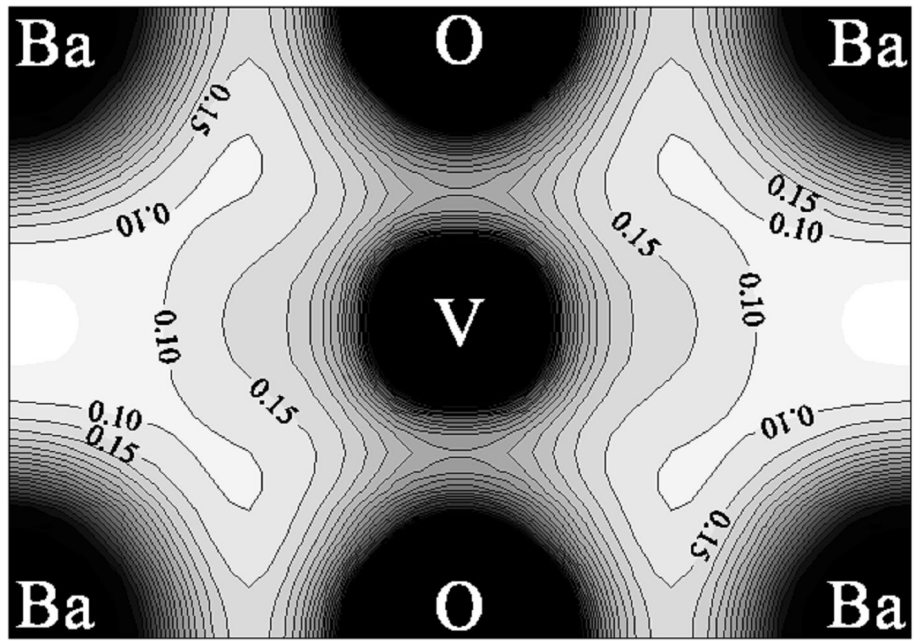

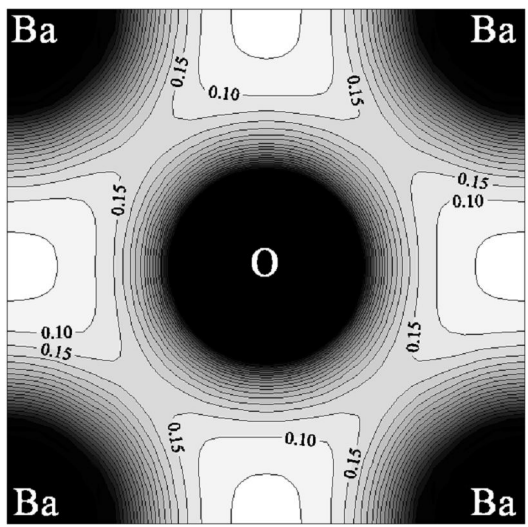 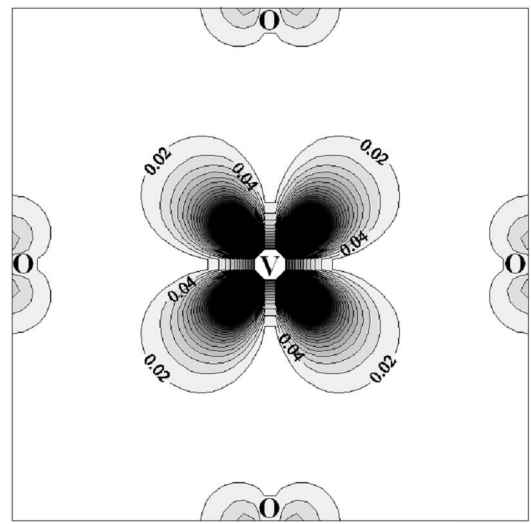

**Fig.3.** The charge density distribution maps in: 1-(110), 2-(100) and 3-(200) planes of BaVO$_3$ perovskite. The partial contribution of occupied conduction band electronic states only is shown in map 3. The interval between contours is 0.05 $e$/Å$^3$ for maps 1-2 and 0.02 $e$/Å$^3$ for map 3.